\documentstyle[aps,prl,graphicx,floats]{revtex}

\newcommand{\beq}{\begin{equation}}
\newcommand{\eeq}{\end{equation}}

\begin{document}
\title{{\bf Universality and scaling study of the critical behavior
            of the two-dimensional Blume-Capel model in short-time dynamics}}
\author{Roberto da Silva{\bf 
\thanks{E-mail: rsilva@dfm.ffclrp.usp.br}, }
        Nelson A. Alves{\bf 
\thanks{E-mail: alves@quark.ffclrp.usp.br}, and }
        J. R. Drugowich de Fel\'{\i}cio{\bf 
\thanks{E-mail: drugo@usp.br}}}
\address{{\it Departamento de F\'{\i}sica e Matem\'{a}tica, 
     FFCLRP Universidade de S\~{a}o Paulo, Avenida Bandeirantes 3900,}\\
CEP 014040-901, \thinspace\ Ribeir\~{a}o Preto, S\~ao Paulo, Brazil}
\date{February, 26, 2002}
\maketitle

\begin{abstract}
In this paper we study the short-time behavior of the Blume-Capel model at 
the tricritical point as well as along the second order critical line. 
Dynamic and static exponents are estimated by exploring scaling relations
for the magnetization and its moments at early stage of the dynamic
evolution.
 Our estimates for the dynamic exponent, at the tricritical point, are 
$z= 2.215(2)$ and $\theta= -0.53(2)$.    
\end{abstract}

\vskip0.1cm {\it Keywords:} short-time dynamics, critical phenomena, 
  dynamic exponent, Blume-Capel model, Monte Carlo simulations.

\vskip0.1cm {\it PACS-No.: 64.60.Fr, 64.60.Ht, 02.70.uu, 75.10.Hk}


\section{INTRODUCTION}

\vspace*{-0.5pt}
\indent
Numerical simulation in the short-time regime has become an important
tool to study phase transitions and critical phenomena. The reason is
that universality and scaling behavior are already present in the dynamic
systems since the early stages of their evolution \cite{Janssen1,Huse}. 
 Moreover, this kind of approach reveals the existence of a new and
unsuspected critical exponent. As shown by Janssen {\it et al.}\cite
{Janssen1} on basis of renormalization group theory, if we tune the
parameters at their critical values but with initial configurations
characterized by nonequilibrium states, the time evolution of quantities
like magnetization exhibit a polynomial behavior
governed by an exponent $\theta$, which is independent of the known set of
static exponents and of the dynamical critical exponent $z$. This new
exponent characterizes the so called ``critical initial slip'', the
anomalous behavior of the magnetization when the system is quenched to the
critical temperature $T_{c}$. Working with systems without conserved
quantities, model A in the terminology of Halperin {\it et al.} \cite{HH1},
Janssen {\it et al.} found a scaling form for the moments of the magnetization,
which sets soon after a microscopic time scale $t_{mic}$. 
Those relations have been
confirmed in several numerical experiments \cite{Li95,Sch97,Review}.
 For the $k{\rm th}$ moment of the magnetization, this scaling form reads 
\begin{equation}
 M^{(k)}(t,\tau ,L,m_{0})= 
 b^{-k\beta /\nu } M^{(k)}(b^{-z}t,b^{1/\nu }
 \tau,b^{-1}L,b^{x_{0}}m_{0}) \, .                        \label{magk}
\end{equation}
Here $b$ is an arbitrary spatial scaling factor, $t$ is the time evolution
and $\tau $ is the reduced temperature, $\tau =(T-T_{c})/T_{c}$. As usual,
the exponents $\beta $ and $\nu $ are the well-known static exponents, whereas
$z$ is the dynamic one. Equation (\ref{magk}) depends on the initial
magnetization $m_{0}$ and gives origin to the new exponent $x_{0}$, the
scaling dimension of the initial magnetization, related to 
$\theta$ by $\theta =(x_{0}-\beta /\nu )/z$.

For a large lattice size $L$ and small initial magnetization $m_0$, the
system in its early stage presents small spatial and temporal correlation
lengths, which may eliminate usual finite size problems. In this limit, if we
choose the scaling factor $b=t^{1/z}$ \cite{Janssen1,Sch97,Review} at the
critical temperature ($\tau =0$), we obtain
\begin{equation}
  M(t,m_{0})\sim m_{0}t^{\theta }                        \label{mag0}
\end{equation}
from the scaling relation (\ref{magk}). 
  The exponent $\theta $ has been calculated for 
the two-dimensional (2d) \cite{Sch97,Grass95} and
  three-dimensional (3d) \cite{Grass95,Li94} Ising models, 
    2d 3-state Potts model \cite{Sch97}, 
Ising model with next-nearest neighbor interactions 
\cite{AiJun} and Ising model with a line of defects \cite{Simoes}. 
 In addition, this short-time universal behavior was
found in irreversible models with synchronous \cite{Tania e Drugo} and
continuous time dynamics \cite{Mendes}. In all of those cases, a positive
value for $\theta $ has been found, that indicates a surprising initial 
increasing of the magnetization in the short-time regime 
$t_{0}\sim m_{0}^{-z/x_{0}}$.
 This effect can be related to a ``mean field'' behavior since
the system presents small correlation length in the beginning of the 
time evolution. 
 Thus, when the system is quenched to the critical temperature 
$T_{c}$, it feels as being in an ordered state since  $T_{c} < T_c^{MF}$
\cite{Zhang99}.

 On the other hand, as shown by Janssen and Oerding \cite{Janssen2}, the
behavior of a thermodynamic system is more complex at a tricritical point
(TP) and the corresponding exponent $\theta$ may attain negative values.

At a tricritical point the magnetization shows a crossover from the
logarithmic behavior $M(t)\sim m_{0}({\rm ln}(t/t_{0}))^{-a}$ at short times
$t<<m_{0}^{-4}$ to $t^{-1/4}$ power law with logarithmic corrections, 
$M(t)\sim (t/{\rm ln}(t/t_{0}))^{-1/4}$ in 3 dimensions.
 The above behavior can be stated in the generalized form 
\begin{equation}
  M(t)=m_{0}\,({\rm ln}(t/t_{0}))^{-a}F_{M}
 \left( \left( \frac{t}{{\rm ln}(t/t_{0})}
 \right)^{1/4}({\rm ln}(t/t_{0}))^{-a}\,m_{0}\right) \,,     \label{mnon}
\end{equation}
where $F_M(x)\sim 1$ or $F_M(x)\sim 1/x$, respectively
for vanishing and large arguments.
 Below 3 dimensions it reduces to the scaling form 
\begin{equation}
 M(t)\sim m_{0}t^{\theta}\,.                              \label{mt}
\end{equation}
Here $\theta$ is the exponent related to the 
tricritical point of the relaxation process at early times
and it is expected to assume negative values.

In this paper, we perform short-time Monte Carlo (MC) simulations to 
explore the critical dynamics of the 2d Blume-Capel model. We evaluate 
the dynamic exponents $\theta$ and $z$, besides the static
exponents $\nu$ and $\beta$ at the tricritical  point. To the best of our 
knowledge this is the first time it is done numerically. We also estimate
 the dynamic exponents along the critical line. We observe a clear trend 
 toward the values of $z$ and $\theta$ for the corresponding 2d Ising values
 when the crystal field $D$ becomes large and negative, indicating
 dynamic universality in the limit 
$D\rightarrow -\infty$. 
  
The paper is organized as follows. In the next section we present the model
and its phase diagram. Sec. III contains the main scaling relations and
describes our short-time MC simulations. Results are presented for
critical points on the second-order line. In Sec. IV, we explore the 
short-time dynamics to study the tricritical behavior. Sec. V contains a brief outlook and concluding remarks.


\section{The model}
\noindent 
 The Blume-Capel \cite{Blume} (BC) model is a spin-1 model
which has been used to describe the behavior of ${\rm ^3He}-{\rm ^4He}$
mixtures along the $\lambda$ line and near the critical mixing point. 
Apart from its practical interest, the BC model has intrinsic interest 
since it is the simplest generalization of the Ising model ($s=1/2$) 
exhibiting a rich phase diagram with first and second-order transition 
lines besides a tricritical point. 
 In real systems tricritical points appear in ${\rm ^3He}-{\rm ^4He}$ 
mixtures, such that when a small fraction of  ${\rm ^3He}$ is added 
to ${\rm ^4He}$, 
a critical line terminates at a concentration of ${\rm ^3He}$ 
approximately at $0.67$.
The BC model or its well known generalization, the
Blume-Emery-Grifftihs model \cite{BEG,Lawrie} was studied by
mean-field approximation, real space and renormalization group schemes 
\cite{RG}, Monte Carlo renormalization group approach \cite{MCRG} and
finite-size scaling combined with conformal invariance 
\cite{Stilck,Deborah,Plascak}. 
 The Hamiltonian of the two-dimensional model is 
\begin{equation}
 H=-J\sum\limits_{<i,j>}S_{i}S_{j}+D\sum\limits_{i=1}S_{i}^{2}\,,  \label{b1}
\end{equation}
where $<i,j>$ indicates nearest neighbors on $L^{2}$ lattices and 
$S_{i}=\{-1,0,1\}$. 
The parameter $J$ is the exchange coupling constant and $D$ is the 
crystal field. We show its phase diagram in Fig. 1. Table I
lists points on the second order critical line and the tricritical point
where we have performed our simulations. 
 Points in Table I were obtained from \cite{Beale}
and from a private communication of the authors in \cite{Plascak}.
 That table also contains our results for the corresponding 
critical and tricritical exponents. 

 We remark that along the critical line, this model presents a critical 
behavior similar to that of the Ising model. However, exactly at the 
tricritical point the exponents change abruptly. They are given by the 
dimensions of the irreducible representations of the Virasoro algebra 
\cite{Friedan,Cardy} with central charge (conformal anomaly number) 
$c = 7/10$ \cite{Deborah}.
In \cite{Stilck} finite-size scaling combined with conformal invariance 
permited to observe a smooth change between Ising-like and 
tricritical behavior.
In finite systems, Ising-like behavior is reached only when 
$D\rightarrow -\infty $. 
 In that limit $\eta /2(2-1/\nu )\rightarrow 0.125$ that is the exact value
for the Ising model. In our short-time simulations the same kind of 
crossover behavior is observed for the dynamic exponents $z$ and 
$\theta$ when we move along the critical line.

\begin{table}[ht]
\renewcommand{\tablename}{Table}
\caption{\baselineskip=0.8cm Critical parameters and exponents for 
     2d Blume-Capel model.}
\begin{center}
\begin{tabular}{cccccccc}\\
\\[-0.3cm]
$D/J$   & $k_{B}T/J$     & $\theta$ & $z$                & $z$ 
                                                         & $1/\nu$ 
                                                         & $1/\nu$ 
                                                         & $\beta$ \\

         &               &        &(Eq.~(\ref{binder})) &(Eq.~(\ref{binder2})) 
                                  &(Eqs.~(\ref{derivate}) and (\ref{binder})) 
                                  &(Eqs.~(\ref{derivate}) and (\ref{binder2})) 
                                  &(Eq.~(\ref{m1}))   \\ 

critical points & &&&&&&  \\
$0$     & 1.6950         & 0.194(3)   & 2.159(6)   & 2.1057(7)    
        & 0.97(2)        & 0.99(2)    & 0.134(2)     \\
-3      & $2.0855$       & 0.193(5)   & 2.156(5)   & 2.1276(5)   
        & 0.99(1)        & 1.00(1)    & 0.125(2)     \\ 
-5      & $2.1855$       & 0.187(5)   & 2.154(4)   & 2.1387(6)   
        & 0.99(3)        & 1.00(3)    & 0.125(4)     \\ 
tricritical point & &&&&&&  \\
$1.9655$& 0.610          & -0.53(2)   & 2.21(2)    & 2.215(2)    
        & 1.864(6)     & 1.86(2)    & 0.0453(2)    \\ 
\end{tabular}
\end{center}
\end{table}


\section{Non-equilibrium short-time dynamics at a critical point}

\noindent In short-time MC simulations critical slowing down can be
neglected. It happens because spatial and time correlation lengths are small
in the early stages of evolution. On the other hand, we need to deal with 
several samples taken over independent initial configurations 
since the systems which are being simulated are far from 
equilibrium. In fact this approach requires calculation of average 
(over samples) 
magnetization and of its moments $M^{(k)}(t)$,
\begin{equation}
M^{(k)}(t) =\frac{1}{N_{s}L^{kd}}
\sum\limits_{j=1}^{N_{s}} \left(
\sum\limits_{i=1}^{L^{d}}\sigma_{ij}(t)\right)^{k} \, ,       \label{samples}
\end{equation}
where $\sigma_{ij}(t)$ denotes the value of spin $i$ of $j$th sample at the 
$t$th MC sweep. Here $N_{s}$ denotes the \ number of samples and $L^{d}$ is
the volume of the system. This kind of simulation is performed $N_{B}$ times
to obtain our final estimates in function of $t$. In this paper, the dynamic 
evolution of the spins $\{\sigma _{i}\}$ is local and updated 
by the heat-bath algorithm.


\subsection{The critical initial slip}
\noindent 
The evolution of the $k$th moment of magnetization in the
initial stage of the dynamic relaxation can be obtained from 
Eq. (\ref{magk}) for large lattice sizes $L$ at $\tau =0$ with 
$b=t^{1/z}$. This yields 
\begin{equation}
M^{(k)}(t,m_{0})=t^{-k\beta /\nu z}M^{(k)}(1,t^{x_{0}/z}m_{0})\,.  \label{c1}
\end{equation}
By expanding the corresponding first moment equation for small $m_{0}$,
we obtain Eq.~(\ref{mag0}) under the condition that $t^{x_{0}/z}m_{0}$ 
is sufficiently small,
which sets a time scale $t_{0}\sim m_{0}^{-z/x_{0}}$ 
\cite{Janssen1,Li95,Review} where that phenomena can be observed.

In Fig. 2a we present our results for the exponent $\theta$
at the critical point $k_{B}T_{c}/J=1.695$ and $D_{c}/J=0$, 
for lattice size $L=80$ and 5 different initial magnetizations $m_0$. 
Our estimates for each $\theta =\theta(m_{0})$ were obtained 
from $N_{B}=5$ independent bins with $N_{S}=10000$,
for $t$ up to 100 sweeps. 
 Figure 2b illustrates the numerical evaluation  of $\theta$ 
for $m_{0}=0.02$ from a log-log plot of the magnetization versus time. 
 The linear fitting in Fig. 2a gives $\theta =0.193(2)$ with
goodness of fit \cite{Press} $Q=0.72$.

Another method has been recently proposed by Tom\'e and de Oliveira 
\cite{Tania} to evaluate $\theta $. It avoids the sharp preparation of 
samples with defined and nonzero magnetization and the delicate 
numerical extrapolation $m_0 \rightarrow 0$. The method is based on the
time correlation function of the total magnetization, 
\begin{equation}
 C(t)=\frac{1}{L^d}\left\langle \sum\limits_{i=1}^{L^d}
\sum\limits_{j=1}^{L^d}\sigma_{i}(t)\sigma_{j}(0)\right\rangle \,.\label{Tome1}
\end{equation}
Starting from random initial configurations
the above correlation behaves as 
$ C(t) \sim t^{\theta}$, which permits us to obtain the exponent 
$\theta$ from a log-log plot of $C(t)$ versus $t$. 
 We obtained $\theta=0.194(3)$ for $k_{B}T_{c}/J=1.695$ and $D_{c}/J=0$
choosing the time interval $[20-150]$
due to the highest value of $Q$ ($Q=0.99$).
This value is in complete agreement with our above 
estimate of the exponent $\theta$ and it is consistent within
error bars with previous results for the 2d Ising model.
 In table I we also present results for $\theta$ at several
points of the critical line.


\subsection{Dynamic critical exponent $z$}
\noindent 
 The observables in short-time analysis are described by different
scaling relations according to the initial magnetizations. In particular, 
the second moment $M^{(2)}(t,L)$ in Eq.~(\ref{samples}), 
\begin{equation}
M^{(2)}= \frac{1}{L^{2d}}\left\langle \sum\limits_{i=1}^{L^d}
         \sigma_{i}^{2}\right\rangle +\frac{1}{L^{2d}}
         \sum\limits_{i\neq j}^{L^d}\left\langle
         \sigma_{i}\sigma_{j}\right\rangle \,,            \label{second}
\end{equation}
with $m_0 = 0$ behaves as $L^{-d}$ since in the short-time evolution 
the spatial correlation length is very small when compared with the
lattice size $L$. Thus, we arrive at \cite{Sch97,Review}
\begin{equation}
 M^{(2)}(t,L)=t^{-2\beta /\nu z}\,M^{(2)}(1,t^{-1/z}L) \sim
 t^{(d-2\beta/\nu)/z}     \,.                                \label{mag2}
\end{equation}
This equation can be used to determine relations involving static critical
exponents and the dynamic exponent $z$ \cite{Review,Foundations}.
 However, a way to evaluate independently the exponent $z$ is through out 
the time-dependent fourth-order Binder cumulant at the critical temperature 
($\tau =0$), 
\begin{equation}
 U_{4}(t,L,m_{0})=1-\frac{M^{(4)}(t,L,m_{0})}
   {3 \left(M^{(2)}(t,L,m_{0})\right)^{2}}\, ,             \label{binder0}
\end{equation}
which obeys the equation
\begin{equation}
 U_{4}(t,L,m_{0}) = U_{4}(b^{-z}t,b^{-1}L,b^{x_0}m_0) \,.   \label{binder}
\end{equation}
 If we set $m_0=0$, we eliminate the dependence on the exponent 
$x_0$ and the exponent $z$ can be evaluated through scaling collapses 
of the generalized cumulant for different lattice sizes \cite{Li95,Li96}.
 To match the Binder cumulants 
$U_{4}(t_{1},L_{1})$ and $U_{4}(t_{2},L_{2})$ obtained from two time series
for lattice sizes $L_{1}$ and $L_{2}$, respectively, with 
$b=L_{2}/L_{1}$ ($L_{2}>L_{1}$), 
we interpolate the series $U_{4}(t,L_{1})$ to obtain 
$\widetilde{U}_{4}(b^{-z}t,L_{1})$. Next, we define the function
\begin{equation}
\chi ^{2}(z)=\frac{1}{t_{f}-t_{i}}\sum\limits_{t=t_{i}}^{t_{f}}\left[
\widetilde{U}_{4}(b^{-z}t,L_{1})-U_{4}(t,L_{2})\right]^{2}\,,\label{cc0}
\end{equation}
where the best estimate for $z$ corresponds to the one which minimizes 
$\chi^{2}(z)$.

In Fig. 3 we show the scaling collapses of the Binder cumulants for
different lattice sizes. We have collapsed the following pairs of lattices
$(L_{1},L_{2})=(10,20),(20,40)$ and $(40,80)$.
 From the largest pair of lattices we obtained $z=2.159(6)$ in the time
interval $[50-1000]$. 
 Our final error estimate is based on 25 different collapses obtained
from $N_B=5$ independent bins for each lattice size.

 Another universal behavior of the dynamic relaxation process also described
by Eq.~(\ref{magk}) can be obtained with the initial condition $m_{0}=1$ 
\cite{Stauffer92,Stauffer93,Sch96}. This condition is related to another
fixed point in the context of renormalization group approach.
 Thus, starting from an initial ordered state one obtains a power law 
decay of the magnetization at the critical temperature, 
\begin{equation}
 M(t)\sim t^{-\beta /\nu z}\,,                             \label{m1}
\end{equation}
when we choose $b^{-z}t=1$ in the limit
of $L\rightarrow \infty $.
 Taking into account this relation, another method has been 
proposed \cite{Review} to estimate the dynamic exponent $z$. 
This approach uses the second cumulant
\begin{equation}
  U_{2}(t,L)=\frac{M^{(2)}(t,L)}{(M(t,L))^{2}}-1  \, ,      \label{binder1}
\end{equation}
which should take the simple form 
$U_2(t,L\rightarrow \infty) \sim  t^{d/z}$.
 The advantage of this procedure is that curves for different lattices lay on the same 
straight line in a log-log plot without any re-scaling in time.
 However, this technique has not been successful 
in at least two well known models: the
two-dimensional $q=3$ Potts model \cite{Review} and the Ising model with
three spin interactions in just one direction \cite{Simoes}. 
 The reason for the above disagreement may be related to the behavior 
of the second term of r.h.s in Eq.~(\ref{second}) when $m_{0}=1$.
 We have proposed \cite{Rsilva} that this behavior could indeed be obtained
working with the ratio
$F_{2}=M^{(2)}/M^{2}$ with different initial conditions for each
moment. As we know the behavior of the
second moment of the magnetization when samples are initially disordered 
$(m_{0}=0)$ and also the time dependence of the magnetization when samples
are initially ordered $(m_{0}=1)$, we easily obtain
\begin{equation}
 F_{2}(t)=\frac{\left. M^{(2)}(t,L)\right|_{m_{0}=0}}{\left.
(M(t,L))^{2}\right| _{m_{0}=1}} \sim t^{d/z}  \,.           \label{binder2}
\end{equation}

 A log-log plot with error bars for the critical point 
$k_B T_c/J =1.695$ and $ D_c/J=0$ is presented in Fig. 4 for $L=160$.
We obtained $z=2.1057(7)$ with $Q=0.99$ in the range $[30-200]$,
which does not agree with the value obtained from Eq.~(\ref{binder}). 
However, as we move away from the tricritical point, 
the values of $z$ obtained (Table 1) with Eq.~(\ref{binder2}) 
show a clear trend toward the expected value of the dynamic exponent 
$z$ $(z=2.1567(7))$ of the 2d Ising model \cite{Rsilva}.
 On the other hand,
the values obtained from the cumulant in Eq.~(\ref{binder}) remain 
essentially the same along the entire critical line. 
It seems the cumulant $U_4$ is 
less sensitive to crossover effects than the mixed method.


\subsection{Static exponents and universality class}
\noindent 
 The exponent $1/\nu z $ can be obtained by differentiating  
$\ln M(t,\tau,m_0)$ with respect to the temperature at $T_{c}$,
\begin{equation}
\left. {\displaystyle{\partial \ln M(t,\tau,L) \over \partial\tau}}
\right|_{\tau=0}   \sim  t^{1/\nu z}  \, ,               \label{derivate}
\end{equation}
if we consider the scaling relation for the magnetization when 
the initial state of samples is ordered ($m_0=1$) \cite{Li96}.

Our results for $\nu z$ were obtained through finite diferences at 
$T_c \pm \delta$ with $\delta=0.001$. They rely on $N_B=5$ independent bins 
with $N_s=5000$ samples each for $L=160$. The time interval $[80-200]$ 
corresponds to the range where the goodness of fit parameter attains 
its highest value $(Q=0.99)$.
 
 In Fig. 5 we show the log-log behavior of the derivative
$\partial_\tau {\rm ln} M(t)$ at $k_B T_c/J= 1.695$ and $D_c/J=0$.
 In Table I we present our final estimates of $1/\nu$. The 6th column
is obtained with the estimates of $z$ from Eq.~(\ref{binder}) 
(data in 4th column),
while the 7th column corresponds to estimates for $\nu$ with values of
$z$ from Eq.~(\ref{binder2})
(data in 5th column).

Since we have already collected estimates for $\nu z$, it is
straightforward to obtain estimates for $\beta$ following Eq.~(\ref{m1}). 
 Our estimates of $\beta$ are presented in the last column.
 Our values in Table I can be compared with theoretical predictions
for an Ising like critical point ($1/\nu =1, \beta =1/8$).


\section{Results from short-time dynamics at the tricritical point}

From the results presented in Refs. \cite{Janssen1} and \cite{Janssen2}
 we can describe 
the time dependence of the magnetization $(k=1)$ for the 2d Blume-Capel model as 
\begin{equation}
M(t)\sim \left\{\begin{array}{ccccc}
m_{0}t^{\theta }~,  & 0<t<t_{0}~, & ~{\rm where}~\theta \geq 0 & \; & \; \\ 
t^{-\beta /\nu z}~, & t_{0}<t<t_{I} & \; & \; & \; \\ 
\; & \; & \; & \; & \; \\ 
m_{0}t^{\theta}~,   & 0<t<t_{0}~, & ~{\rm where}~\theta \leq 0 & \; & \; \\ 
t^{-\beta /\nu z}~, & t_{0}<t<t_{I} & \; & \; & \;
\end{array}
\begin{array}{c}
\text{critical} \\ 
\text{point} \\ 
\\ 
\text{tricritical} \\ 
\text{point}
\end{array}
\right. 
\end{equation}
for an initial small magnetization $m_0$. Here $t_I$ stands for
the time before the system has reached the thermalization.
 
 We also included in Table I our estimates for $\theta, z, 1/\nu$ and 
$\beta$ at the tricritical point
$k_{B}T_{t}/J=0.610$ and $D_{t}/J=1.9655$ working with lattice size $L=80$.

 In Fig. 6a we show the values of $\theta$ for 5 different initial 
magnetizations $m_0$ at the tricritical point and in Fig. 6b the time
evolution of the magnetization for $m_{0}=0.08$ in a log-log plot.
 Our estimates for each  $\theta (m_{0})$ were obtained from $N_{B}=20$ 
independent bins with $N_{S}=10000$ samples, for $t$ up to 80 sweeps. 
 The linear extrapolation in Fig. 6a gives $\theta=-0.53(2)$ 
with goodness of fit $Q=0.75$.
 The corresponding study with the time correlation function
in Eq.~(\ref{Tome1}) also gives  $\theta=-0.53(2)$ with $Q=0.99$ in the
time interval $[20-80]$.
 The generalization of the dynamic scaling relation for the $\it k$th
moment of the magnetization at a tricritical point can be written as
 \cite{Bonfim2} 

\begin{equation}
  M^{(k)}(t,\tau ,g,L,m_{0})=
  b^{-k\beta_{t} / \nu}
M^{(k)}(b^{-z}t,b^{1/\nu}\tau,b^{\phi_{t}/\nu}g,b^{-1}L,b^{x_0}m_{0}).
                                                              \label{magtri}
\end{equation}
 It differs from the critical case by the scaling field 
$g$ that measures
the shunting line of critical point along of tangent transition line (at
the tricritical point). The quantity $\phi_{t}$ is the known as the 
crossover exponent. At the criticality we have $\phi_t = g = 0$.

 We show in Fig.7 scaling collapses of the cumulant $U_4(t,L)$ at the
tricritical point, whose behavior is quite different of the corresponding
one at a critical point (Fig. 3).
 Our first estimate based on Eq.~(\ref{binder}) leads to $z=2.21(2)$. 
This value was
 obtained from the pair of largest lattice sizes $L=40$ and $80$.
 The value remains the same when we consider the time interval
$[10, 1000]$ or $[200, 1000]$.
 The second estimate for the dynamic exponent, based on 
Eq.~(\ref{binder2}) gives $z=2.215(2)$. It was obtained from 
a larger lattice  ($L=160$) in the time interval $[30, 200]$,
with $Q=0.71$. 
 We do not show the log-log plot of $F_2(t)$ in this case because 
it is quite similar  to Fig. 4.
 We had to restrict the time interval, when compared with the $U_4$
calculation, in order to obtain acceptable values for $Q$. 
 Here, in contrast to the estimates of $z$ (4th and 5th 
column) at the critical points listed in Table I, both methods
lead to the same numerical result.

The evaluation of static exponents $\nu$ and $\beta$ at the tricritical 
point follows the same procedure as applied to the critical line. 
 Estimates exhibited in Table I are in good agreement with results 
provided by conformal invariance $\nu=9/5$ and $\beta=1/24$.

Next, in order to study the influence of the local dynamics over the 
values of the exponents we recall the simulations with Glauber dynamics
performed by de Alcantara Bonfim.
 Our estimates of $\beta/\nu z$ obtained from the decay of the magnetization 
(starting from initialy ordered state) is
$0.0381(1)$ with the heat-bath dynamics while $\beta/\nu z =0.0377(7)$
when updated following Glauber rules.



\section{CONCLUSIONS}

\indent

	We have performed short-time Monte Carlo simulations to evaluate 
dynamic and static exponents at critical and tricritical points of 
the spin-1 Blume-Capel model. According to analytical predictions 
by Janssen and Oerding, a negative value for the new exponent $\theta$ 
was obtained for the tricritical point. 
The dynamic exponent $z$ was estimated by colapsing the fourth-order 
Binder cumulant and also by the function $F_2(t)$ which explores time 
evolution of the magnetization starting from different initial conditions. 
The value of $z$ was used to obtain the static exponents 
$\beta$ and $\nu$ which resulted in good agreement with the exact results 
provided by the tricritical Ising model \cite{Deborah,Friedan}.
	The dynamic exponents also were calculated along the critical line. 
Estimates for $\theta$ are in good agreement with known result for the 
2d Ising model. However, near the tricritical point we found 
strong crossover effects on dynamical exponent $z$ when using the 
recently proposed \cite{Rsilva} technique based on mixed initial conditions. 
 On the other hand, the fourth-order Binder cumulant is almost free from 
crossover effects.


\newpage \vspace{0.4cm}

{\bf Acknowledgments} \vspace{0.4cm}

R. da Silva and J. R. Drugowich de Fel\'{\i}cio gratefully acknowledge 
support by FAPESP (Brazil)
 and N.A. Alves by CNPq (Brazil).  
 The authors deeply thank also to J. A. Plascak and J. C. Xavier for their 
communication about the values of critical points for $D<0$ 
and DFMA (IFUSP) for the computer facilities extended to us.



\newpage \cleardoublepage

\begin{figure}[b]
\begin{center}
\begin{minipage}[t]{0.95\textwidth}
\centering
\includegraphics[width=0.72\textwidth]{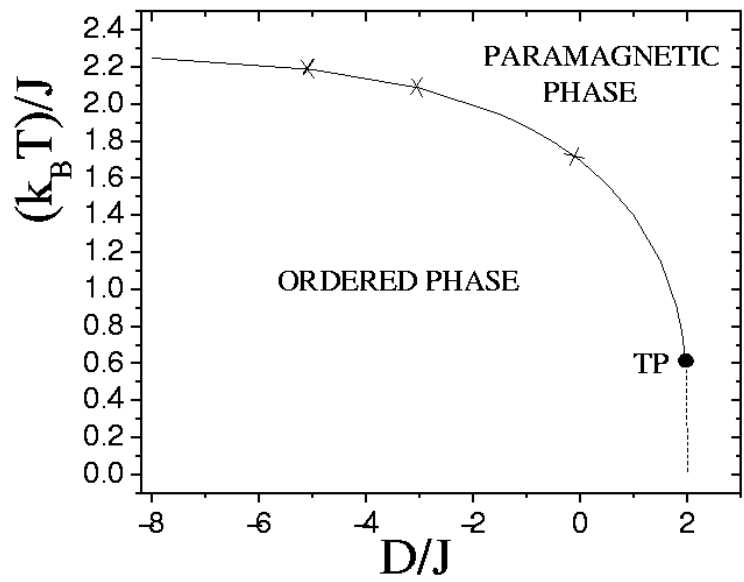}
\renewcommand{\figurename}{(Fig.1)}
\caption{Phase diagram of the Blume-Capel model. The dashed curve
         is a first-order transition line and the solid is a second order
         one.
         These curves are connected by a tricritical point (TP). 
         The marked points $(\times,\bullet)$ correspond to the simulated
         values.}
\label{Fig. 1}
\end{minipage}
\end{center}
\end{figure}

\newpage \cleardoublepage

\begin{figure}[b]
\begin{center}
\begin{minipage}[t]{0.95\textwidth}
\centering
\includegraphics[width=0.72\textwidth]{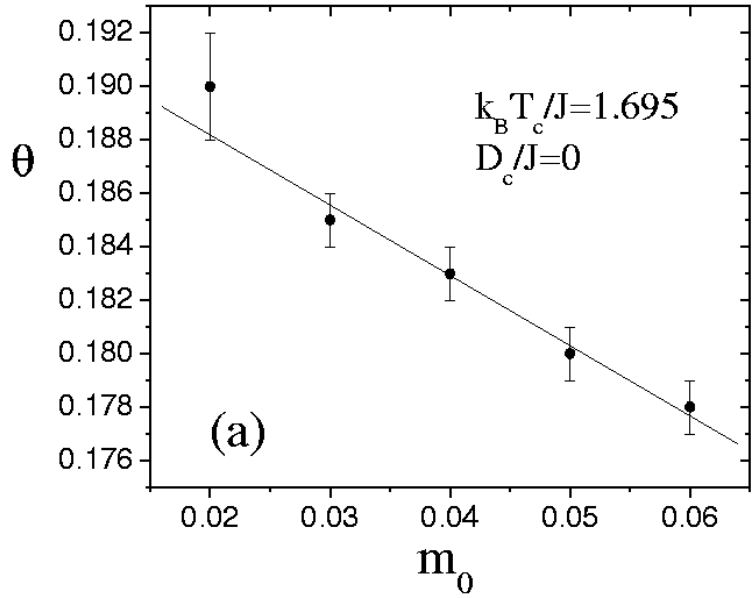}
\renewcommand{\figurename}{(Fig.2a)}
\caption{Exponent $\theta$ in function of initial magnetizations $m_0$
         for square lattices with $L=80$. The straight line 
         is a least-square fit to the data.}
\label{Fig. 2}
\end{minipage}
\end{center}
\end{figure}

\begin{figure}[b]
\begin{center}
\begin{minipage}[t]{0.95\textwidth}
\centering
\includegraphics[width=0.72\textwidth]{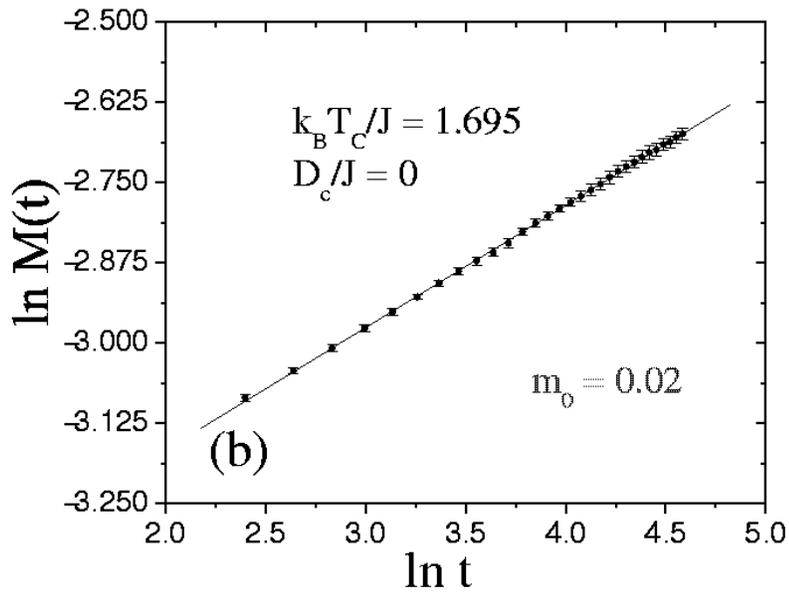}
\renewcommand{\figurename}{(Fig.2b)}
\caption{Time evolution of the magnetization for $L=80$ and $m_0=0.02$.}
\label{Fig. 2b}
\end{minipage}
\end{center}
\end{figure}

\newpage \cleardoublepage

\begin{figure}[b]
\begin{center}
\begin{minipage}[t]{0.95\textwidth}
\centering
\includegraphics[width=0.72\textwidth]{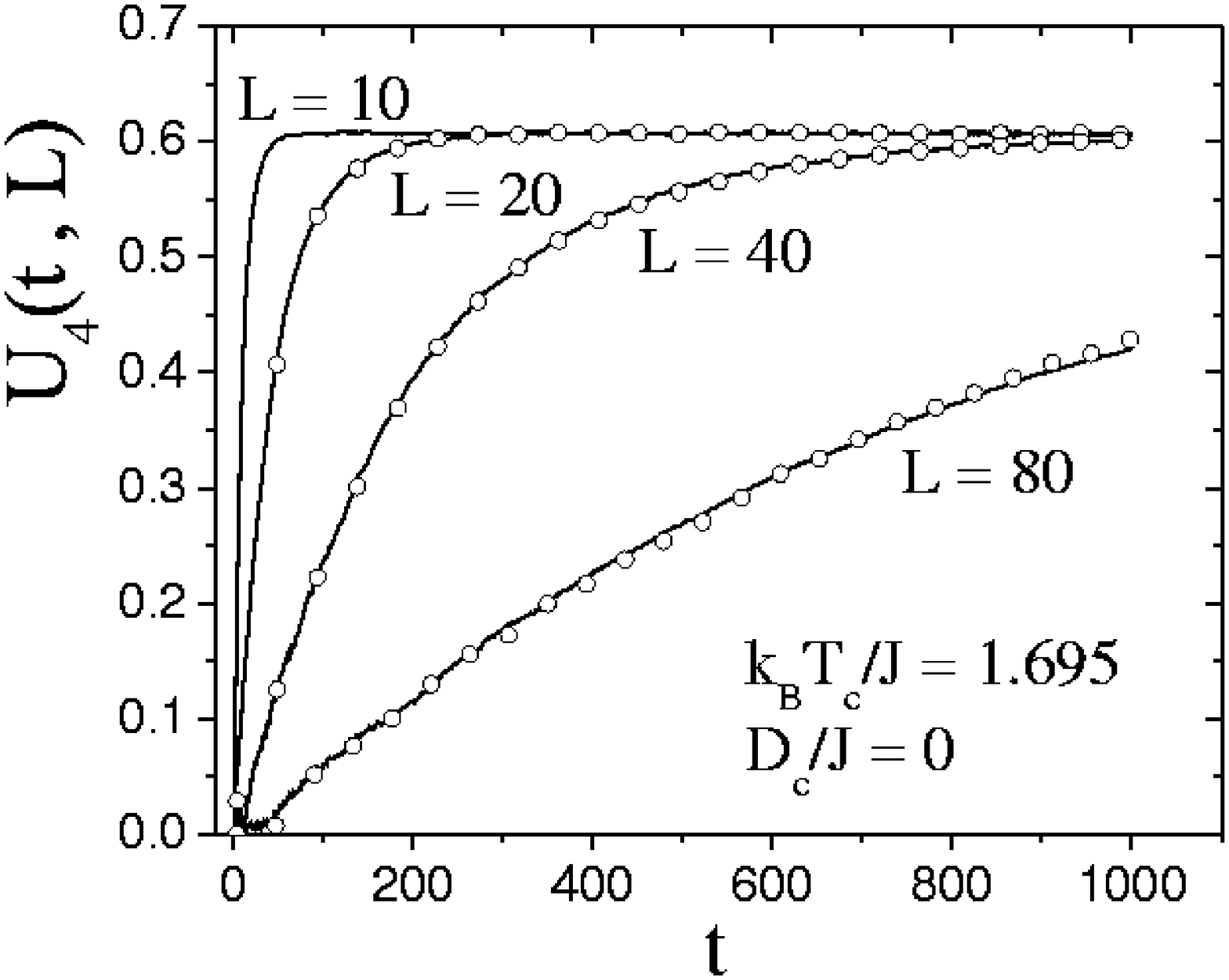}
\renewcommand{\figurename}{(Fig.3)}
\caption{Cumulants $U_4(t,L)$ for $L=10,20, 40$ and $80$ for initial
         magnetization $m_0=0$. The dots on the lines show the 
         cumulants for lattice sizes $L/2$ rescaled in time with $z$ given by 
         Eq.~(\ref{cc0}).}
\label{Fig. 3}
\end{minipage}
\end{center}
\end{figure}

\begin{figure}[b]
\begin{center}
\begin{minipage}[t]{0.95\textwidth}
\centering
\includegraphics[width=0.72\textwidth]{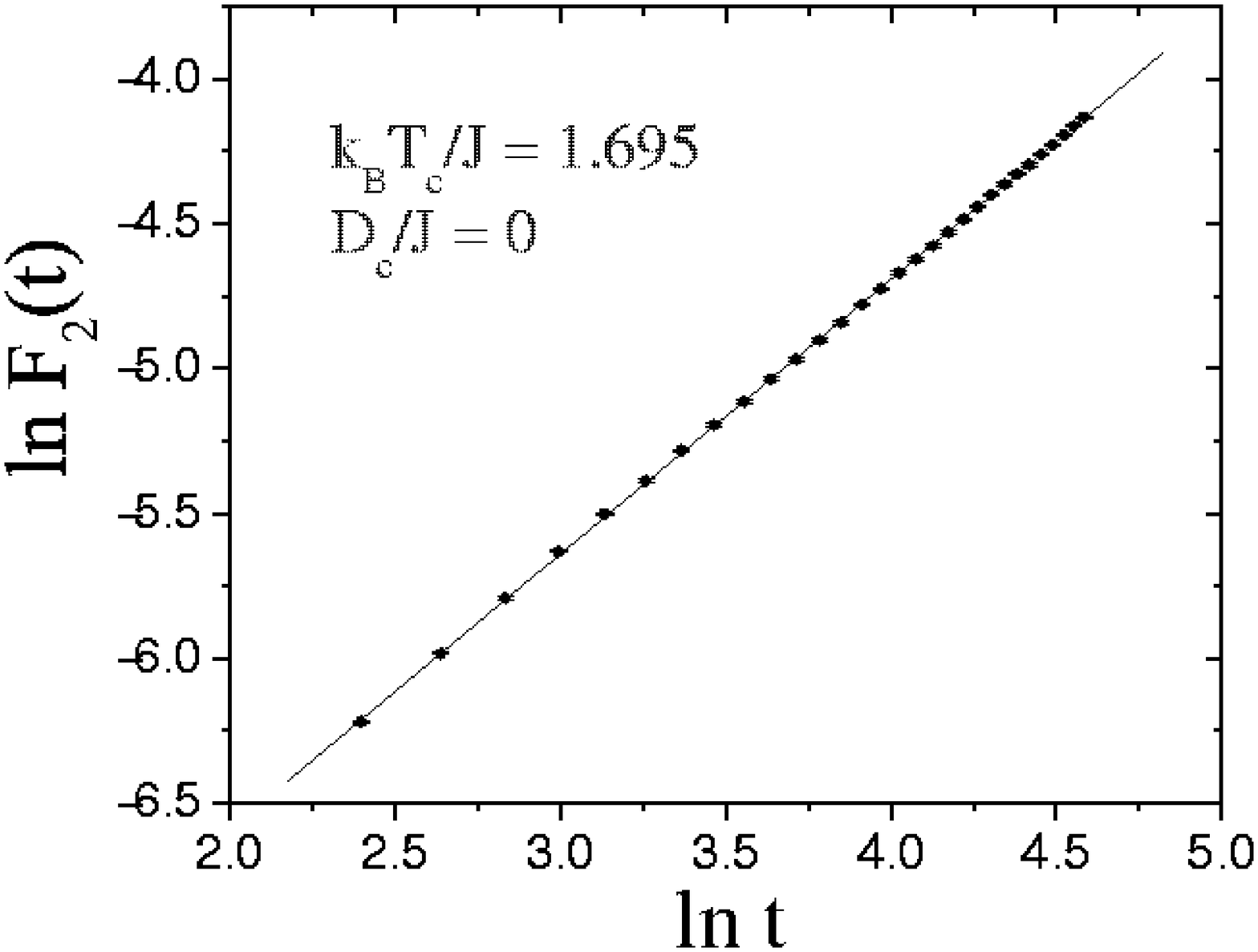}
\renewcommand{\figurename}{(Fig.4)}
\caption{Time evolution of $F_2(t)$ for $L=160$ with mixed initial
         magnetizations.}
\label{Fig. 4}
\end{minipage}
\end{center}
\end{figure}

\newpage \cleardoublepage

\begin{figure}[b]
\begin{center}
\begin{minipage}[t]{0.95\textwidth}
\centering
\includegraphics[width=0.72\textwidth]{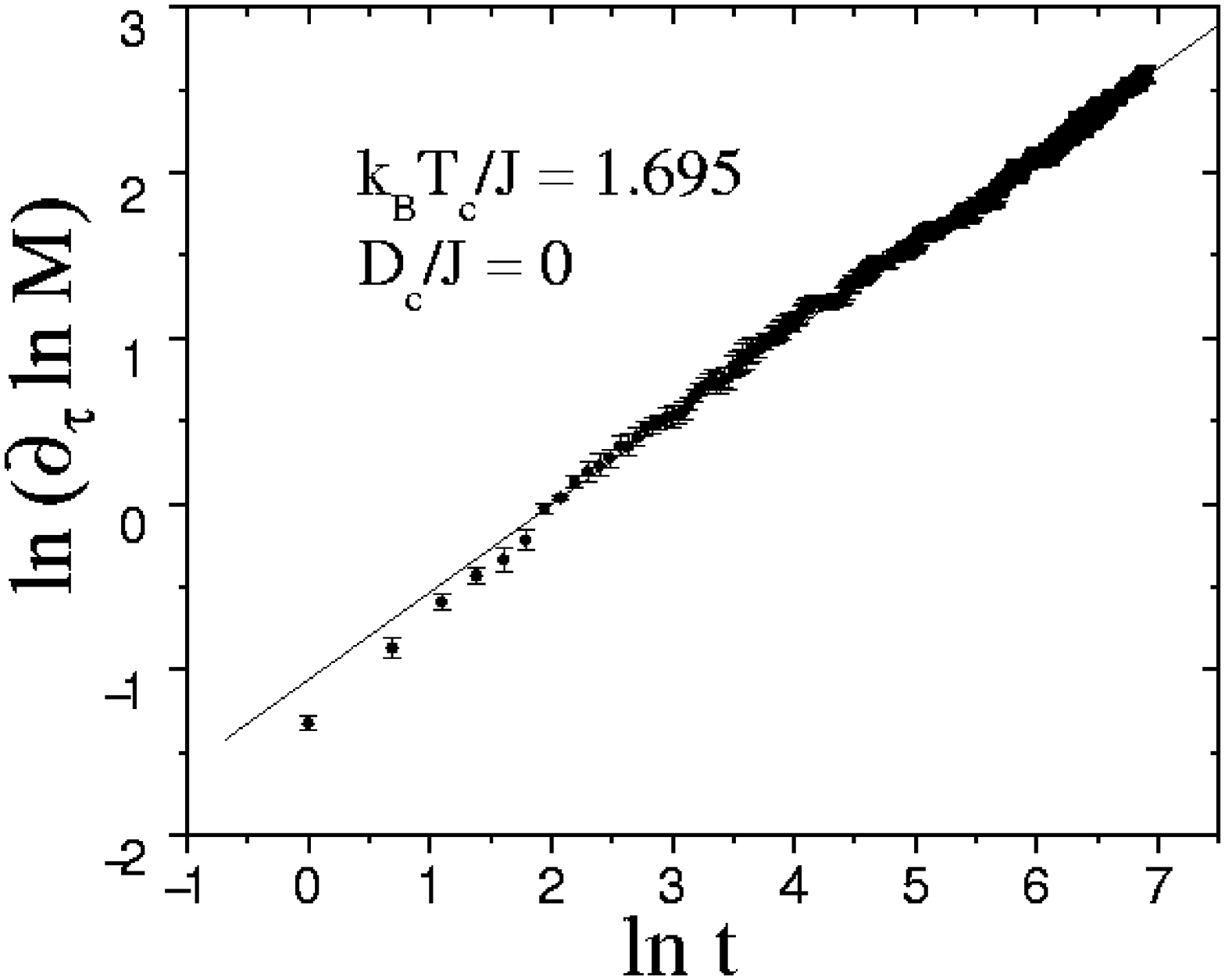}
\renewcommand{\figurename}{(Fig.5)}
\caption{Time evolution of the derivative 
         $\partial_{\tau}{\rm ln}M(t,\tau)|_{\tau=0}$ for $L=160$
         and initial magnetization $m_0=1$. }
\label{Fig. 5}
\end{minipage}
\end{center}
\end{figure}

\begin{figure}[b]
\begin{center}
\begin{minipage}[t]{0.95\textwidth}
\centering
\includegraphics[width=0.72\textwidth]{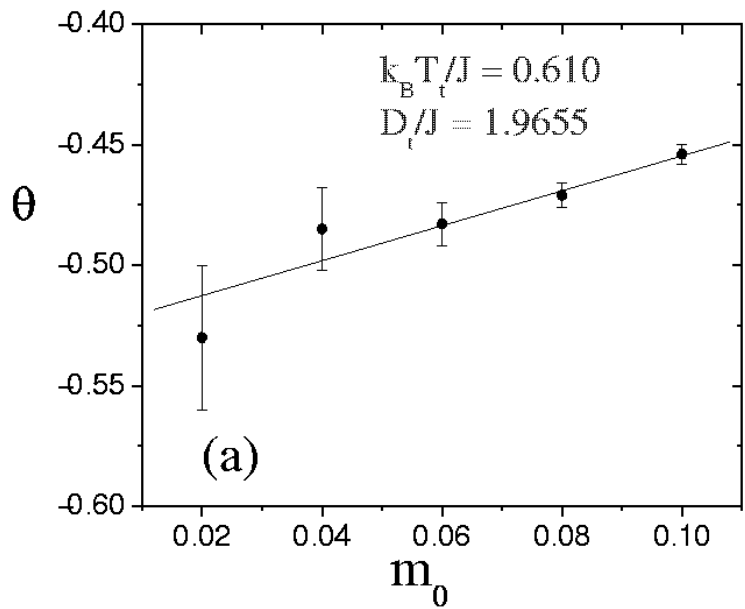}
\renewcommand{\figurename}{(Fig.6a)}
\caption{Exponent $\theta$ in function of initial magnetizations $m_0$
         for $L=80$ at the tricritical point. 
         The straight line is a least-square fit to the data.}
\label{Fig. 6}
\end{minipage}
\end{center}
\end{figure}

\newpage \cleardoublepage

\begin{figure}[b]
\begin{center}
\begin{minipage}[t]{0.95\textwidth}
\centering
\includegraphics[width=0.72\textwidth]{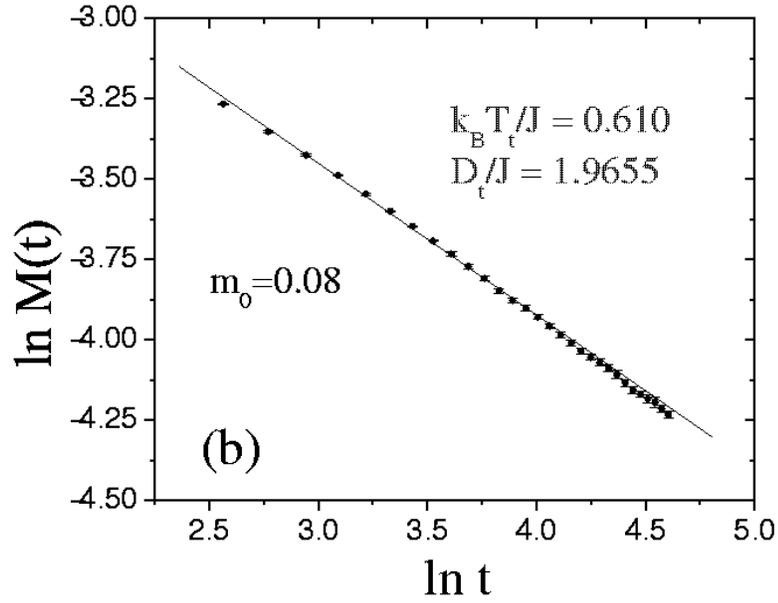}
\renewcommand{\figurename}{(Fig.6b)}
\caption{Time evolution of the magnetization for $L=80$ and fixed $m_0=0.08$
         at the tricritial point.}
\label{Fig. 6b}
\end{minipage}
\end{center}
\end{figure}

\begin{figure}[b]
\begin{center}
\begin{minipage}[t]{0.95\textwidth}
\centering
\includegraphics[width=0.72\textwidth]{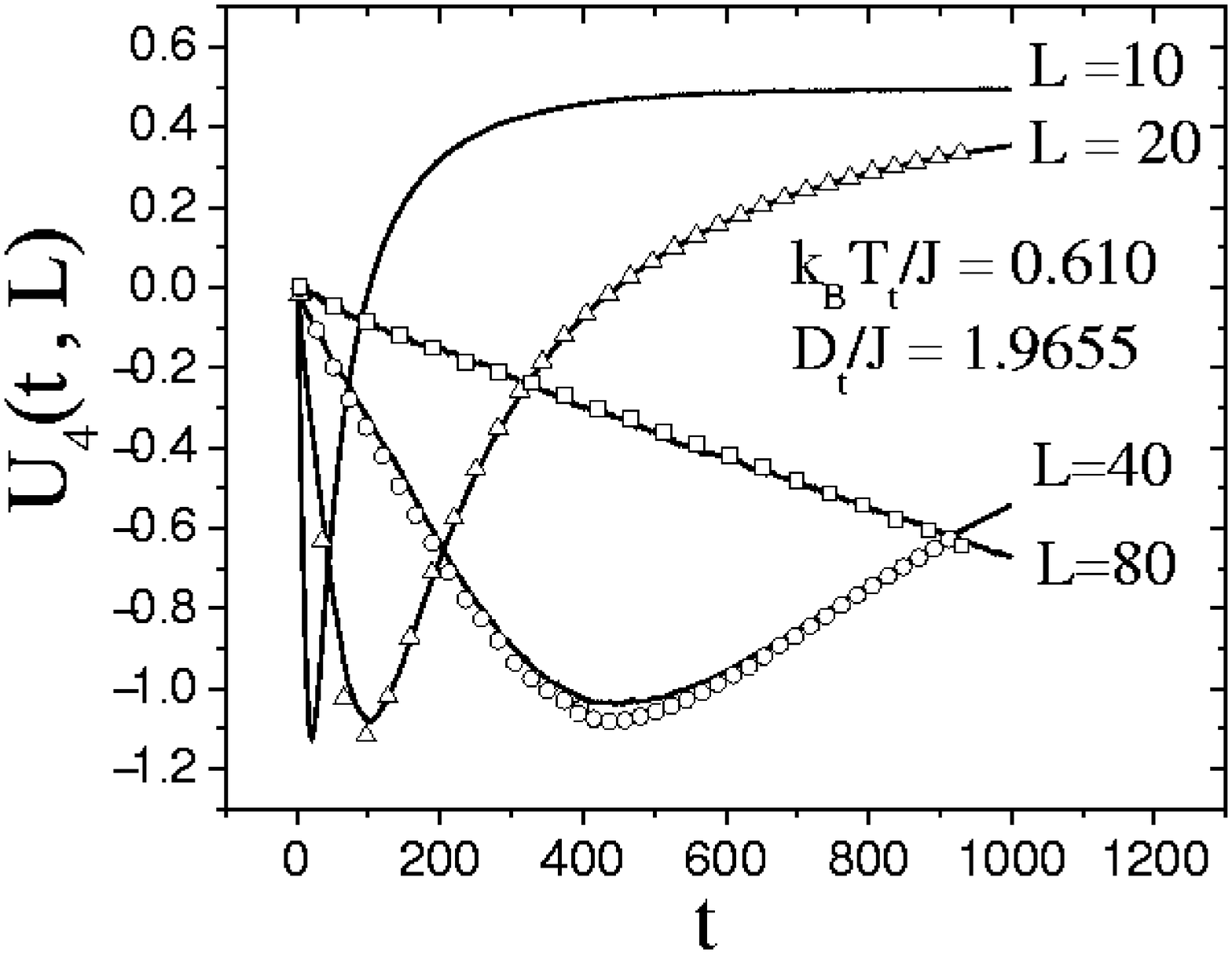}
\renewcommand{\figurename}{(Fig.7)}
\caption{Cumulants $U_4(t,L)$ for $L=10,20, 40$ and $80$ for an initial
         magnetization $m_0=0$ at the tricritical point.
         The dots on the lines show the 
         cumulants with lattice sizes $L/2$ rescaled in time with $z$ given by 
         Eq.~(\ref{cc0}).}
\label{Fig. 7}
\end{minipage}
\end{center}
\end{figure}


\begin{references}

\bibitem{Janssen1}  H. K. Janssen, B. Schaub and B. Schmittmann, 
                    Z. Phys. B {\bf 73}, 539 (1989).

\bibitem{Huse}  D. A. Huse, Phys. Rev. B {\bf 40}, 304 (1989).

\bibitem{HH1}  B. I. Halperin, P. C. Hohenberg and S-K. Ma, 
               Phys. Rev. B {\bf 10}, 139 (1974).

\bibitem{Li95}  Z. B. Li, L. Sch\"{u}lke and B. Zheng, 
                Phys. Rev. Lett. {\bf 74}, 3396 (1995).

\bibitem{Sch97}  K. Okano, L. Sch\"{u}lke, K. Yamagishi and 
                 B. Zheng, Nucl. Phys. B {\bf 485} [FS], 727 (1997).

\bibitem{Review}  B. Zheng, Int. J. Mod. Phys. B {\bf 12}, 1419 (1998).

\bibitem{Grass95} P. Grassberger, Phys. A {\bf 214}, 547 (1995).

\bibitem{Li94}  Z.-B. Li, U. Ritschel, B. Zheng, 
                J. Phys. A: Math. Gen. {\bf 27}, L837 (1994).

\bibitem{AiJun} Y. E. AiJun, Pan ZhiGang, Chen Yuan and Li ZhiBing,
                Commun. Theor. Phys. (Beijing, China) {\bf 33}, 205 (2000). 

\bibitem{Simoes} C. S. Sim\~{o}es and J. R. Drugowich de Fel\'{\i}cio, 
                 J. Phys. A {\bf 31}, 7265 (1998);
                 T. Tom\'e, C. S. Sim\~{o}es and J. R. Drugowich 
                 de Fel\'{\i}cio, 
                 Mod. Phys. Lett. B {\bf 15}, 487 (2001);
                 L. Wang, J. B. Zhang, H. P. Ying and D. R. Ji,
                 Mod. Phys. Lett. B {\bf 13}, 1011 (1999).

\bibitem{Tania e Drugo} T. Tom\'e and J. R. Drugowich de Fel\'{\i}cio, 
                 Mod. Phys. Lett. B {\bf 12}, 873 (1998).

\bibitem{Mendes} J. F. F. Mendes and M. A. Santos, Phys. 
                 Rev. E {\bf 57}, 108 (1998).

\bibitem{Zhang99} J.-B. Zhang, L. Wang, D.-W. Gu, H.-P. Ying and D.-R. Ji,
            Phys. Lett. A {\bf 262}, 226 (1999), and references therein.

\bibitem{Janssen2} H. K. Janssen and K. Oerding, 
                   J. Phys. A: Math. Gen. {\bf 27}, 715 (1994).

\bibitem{Blume} M. Blume, Phys. Rev. {\bf 141}, 517 (1966).
                H. W. Capel, Physica {\bf 32}, 966 (1966);
                                    {\bf 33}, 295 (1967);
                                    {\bf 37}, 423 (1967).

\bibitem{BEG}  M. Blume, V. J. Emery and R. B. Griffiths, 
               Phys. Rev. A {\bf 4}, 1071 (1971).

\bibitem{Lawrie}  I. D. Lawrie and S. Sarbach, in 
       {\it Phase Transitions and Critical Phenomena}, vol.9. 
       Edited by C. Domb and J.L. Lebowitz, 1984 (Academic Press).

\bibitem{RG}  A. N. Berker and M. Wortis, Phys. Rev. B {\bf 14}, 4969 (1976); 
              T. W. Burkhardt, ibid. {\bf 14}, 1196 (1976).

\bibitem{MCRG}  D. P. Landau and R. H. Swendsen, 
                Phys. Rev. Lett. {\bf 46}, 1437 (1981).

\bibitem{Stilck} F. C. Alcaraz, J. R. Drugowich de Fel\'{\i}cio,
           R. K\"{o}berle and J. F. Stilck, Phys. Rev. B {\bf 32}, 7469 (1985).

\bibitem{Deborah} D. B. Balb\~ao and J. R. Drugowich de Fel\'{\i}cio, 
                  J. Phys. A {\bf 20}, L207 (1987).

\bibitem{Plascak} J. C. Xavier, F. C. Alcaraz, D. Pen\~a Lara and 
                  J. A. Plascak, 
                  Phys. Rev. B {\bf 57}, 11575 (1998).

\bibitem{Beale}  P.D. Beale, Phys. Rev. B {\bf 33}, 1717 (1986).

\bibitem{Friedan}  D. Friedan, Z. Qiu and S. Shenker, 
                   Phys. Rev. Lett. {\bf 52}, 1575 (1984).

\bibitem{Cardy} J. L. Cardy, Nucl. Phys. B {\bf 270}, 186 (1986); 
                                         B {\bf 275}, 200 (1986).

\bibitem{Press}  W. Press {\it et al.}, {\it Numerical Recipes} 
                 (Cambridge University Press, London, 1986).

\bibitem{Tania}  T. Tom\'{e} and M. J. de Oliveira, 
                 Phys. Rev. E {\bf 58}, 4242 (1998).

\bibitem{Foundations}  K. Okano, L. Sch\"{u}lke and B. Zheng, 
      Foundations of Physics {\bf 27}, 1739 (1997).

\bibitem{Li96}  Z. Li, L. Sch\"{u}lke, B. Zheng, 
                Phys. Rev. E {\bf 53}, 2940 (1996).

\bibitem{Stauffer92}  D. Stauffer, Physica A {\bf 186}, 197 (1992).

\bibitem{Stauffer93}  C. M\"{u}nkel, D.W. Heermann, J. Adler, 
   M. Gofman and D. Stauffer, Physica A {\bf 193}, 540 (1993).

\bibitem{Sch96}  L. Sch\"{u}lke, B. Zheng, 
                 Phys. Lett. A {\bf 215}, 2940 (1996).

\bibitem{Rsilva}  R. da Silva, N. A. Alves and J. R. Drugowich de
                  Fel\'{\i}cio, cond-mat/0111288.

\bibitem{Bonfim2} O. F. de Alcantara Bonfim, 
                  J. Stat. Phys. {\bf 48}, 919 (1987).

\end{references}
\end{document}